# A Study on the Impact of Human Resource Accounting on Firm's Value with Respect to Companies Listed in National Stock Exchange

Anil S

*Student, M.Com in Accounting and Taxation*
*M.S. Ramaiah University of Applied Sciences*
*Email- anilkashyap19978@gmail.com*

Sudharani R

*Assistant Professor*
*M.S. Ramaiah University of Applied Sciences*
*Organization Name, City, State, Country*
*Email- sudharani.co.mc@msruas.ac.in*

Dr. Suresh N

*Professor*
*M.S Ramaiah University of Applied Sciences*
*Email- nsuresh.ms.mc@msruas.ac.in*

**Abstract-** The study focuses on the Impact of Employment Benefit Cots on Profitability of Companies listed in national Stock Exchange. The study has considered the Amount spent on Employment Benefit Cots as Independent variable and Profit after tax, Total Assets, Return on Equity, and Return on Asset and Debt equity Ration as Dependent variable. The present study is to analyses the relationship between Employment Benefit Cots and Profit after tax, Total Assets, Return on Equity, Return on Asset and Debt equity Ration. The data is collected of 20 companies listed in National Stock Exchange of 10 years from Annual report of companies. The data collected were analyzed using Panel data Regression in E-Views. Results revealed that there is a significant Relationship between Employment Benefit Cots and Profit after tax, Total Assets, Return on Equity, Return on Asset and Debt equity Ration. The study shows that Employment Benefit Cots impact positively on Firms profitability.

1.   INTRODUCTION

Human resource accounting (HRA) is the process of identifying and reporting investments made in the Human Resources of an organization that are presently unaccounted for in the conventional accounting practice. It is an extension of standard accounting principles. Measuring the value of the human resources can assist organizations in accurately documenting their assets. In other words, human resource accounting is a process of measuring the cost incurred by the organization to recruit, select, train, and develop human assets. Human resources are considered as important assets and are different from the physical assets. Physical assets do not have feelings and emotions, whereas human assets are subjected to various types of feelings and emotions. In the same way, unlike physical assets human assets never gets depreciated. Therefore, the valuations of human resources along with other assets are also required in order to find out the total cost of an organization. In 1960s, Rensis Likert along with other social researchers made an attempt to define the concept of human resource accounting (HRA).

## 1.1 Need for Human Resource Accounting (HRA)

The need for human asset valuation arose as a result of growing concern for human relations management in the industry. Behavioral scientists concerned with management of organizations pointed out the following reasons for HRA

- Under conventional accounting, no information is made available about the human resources employed in an organization, and without people the financial and physical resources cannot be operationally effective.

- The expenses related to the human organization are charged to current revenue instead of being treated as investments, to be amortized over a period of time, with the result that magnitude of net income is significantly distorted. This makes the assessment of firm and inter-firm comparison difficult.

- The productivity and profitability of a firm largely depends on the contribution of human assets. Two firms having identical physical assets and operating in the same market may have different returns due to differences in human assets. If the value of human assets is ignored, the total valuation of the firm becomes difficult.

- If the value of human resources is not duly reported in profit and loss account and balance sheet, the important act of management on human assets cannot be perceived.

- Expenses on recruitment, training, etc. are treated as expenses and written off against revenue under conventional accounting. All expenses on human resources are to be treated as investments, since the benefits are accrued over a period of time.

## 2. LITERATURE REVIEW

Vaddadi, Surarchith and Subhashin (2018) The study aims at ascertaining the relationship between human resource accounting and performance of firm. The study was carried out in ten branches of Indian Nationalized bank located in Andhra Pradesh, India. Quantitative method was used to examine the present study. Researchers identified three independent factors under Human Resource Accounting (HRA) and one dependent factor as performance of a firm. According to findings of the study, shelter cost and training & development cost were strongly correlated with firm's performance, but health and safety cost was moderately correlated with firm's performance. The research helps the banks to identify the importance of investment on human capital.

Dhar, Mutalib and Sobhani (2017) the aim of this paper is to investigate the impact of Human Resource Accounting on organizational performance and to develop a framework that could be beneficial for the researchers, policy makers and investor's community. This study uses a systematic review of literatures that focus on the impact of the factors that influence human resource accounting practice on organizational performance. The findings of the study have been merged together in a proposed framework with the help of the disclosure of human resource accounting and use of intellectual capital accounting on organizational performance with the help of management support and employees' performance.

Chathurika and Silva (2019) This aim of the study is to investigating the impact of expenditure on human capital and human capital accounting on financial performance of Sri Lankan Listed companies. This study employs panel data over five years from 2011 to 2015 of forty listed companies across all sectors in the Colombo Stock Exchange Sri Lanka. The finding of the research implies that spending more on human capital of the organization improves the financial performance. Thus, spending on human resources is no longer a burden to the firm's financial performance.

Dhanabhakyam and Mufliha (2016) The aim of the study is to examines the impact of human resource accounting system on the decision-making areas of human resource management practices. The study data were collected among the 100 respondents from the Head office of State Bank of India and Canara bank in the Kerala region targeted at the staff of human resource department, accounting section and audit control department. Hence the collected data were analyzed using Correlation analysis and Multiple Regression analysis model validated through

ANOVA and F Ratio. The study result was found that all the aspects related to the implementation of Human Resource Accounting is closely associated and has 67.1 percent impact on the decision-making areas of human resource management practices.

Odum and Aroh (2017) This study aims to investigate the effect of Human Capital Accounting Information on the Market Value of firms, with particular reference to the consumables sector of quoted manufacturing firms on the Nigerian Stock Exchange. The study relied primarily on secondary data as the main source of data. Descriptive statistics was used to describe and summarize the behavior of the variables in the study. From the results of analysis only managerial disclosure and health and security costs were found to have a negative effect on our sampled companies' Earnings Per Share (EPS)

### 3. PROBLEM FORMULATION

In the existing literature research on the impact of Human Resource Accounting on firm's value very limited number of variables are considered and most of the researches are being conducted on studying the relationship of human resource accounting and Profitability of companies using excel spread sheet, descriptive statistics, SPSS and other statistical tools and none of the researcher have used Panel data regression for studying the same.

This paper will therefore fill the gap by taking up a comprehensive study on profitability factors of NSE listed companies with specific variables such as total assets, profits after tax, debt equity ratio, and return on equity and return on assets. This research has been conducted using different statistical tools which are not used before to study and analyses the relationship between Human Resource Accounting and Profitability for the period of ten years. The current study also focus on analyzing large number of companies data compared to earlier studies.

### 4. OBJECTIVES

- To study the impact of disclosure of human resource accounting on profitability of companies
- To analyze the impact of Employment benefit cost on Profit after tax and Total asset
- To analyze the impact of Employment benefit cost on Debt Equity Ratio and Return on Equity and return on asset
- To provide suitable suggestions based on study finding

### 5. PROBLEM SOLVING

**5.1 Data Collection and sample size**

For this study variables such as employment benefit cost, Profit after Tax, Return on Assets, Return on Equity, Total Assets and Debt Equity Ratio of 20 NSE listed companies for the period of 10 years ( 2010 to 2019) is considered. The required data for this study is collected from annual reports of the selected companies available in respective company's website and few of the data were collected from money control website.

**5.2 HYPOTHESIS**

H0- There is no significant relationship between Employment Benefit cost and Profit After Tax, Total Assets, Debt Equity Ratio, return on equity, Return on Assets.

H1-There is a significant relationship between Employment Benefit cost and Profit After Tax, Total Assets, Debt Equity Ratio, return on equity, Return on Assets.

## 5.3 Results and discussion

**Results of Fixed effect model of Employment benefit cost and profit after tax**

Dependent Variable: X1PAT
Method: Panel Least Squares
Date: 03/11/20   Time: 11:10
Sample: 2010 2019
Periods included: 10
Cross-sections included: 20
Total panel (balanced) observations: 200

| Variable | Coefficient | Std. Error | t-Statistic | Prob. |
|---|---|---|---|---|
| C | 4140.810 | 329.2926 | 12.57486 | 0.0000 |
| YEBC | 0.356183 | 0.046019 | 7.739888 | 0.0000 |

Effects Specification

Cross-section fixed (dummy variables)

| | | | |
|---|---|---|---|
| R-squared | 0.776998 | Mean dependent var | 5890.529 |
| Adjusted R-squared | 0.752081 | S.D. dependent var | 6800.554 |
| S.E. of regression | 3386.093 | Akaike info criterion | 19.19181 |
| Sum squared resid | 2.05E+09 | Schwarz criterion | 19.53813 |
| Log likelihood | -1898.181 | Hannan-Quinn criter. | 19.33196 |
| F-statistic | 31.18412 | Durbin-Watson stat | 0.370544 |
| Prob(F-statistic) | 0.000000 | | |

**Results of Random effect model of Employment benefit cost and profit after tax**

Dependent Variable: X1PAT
Method: Panel EGLS (Cross-section random effects)
Date: 03/12/20   Time: 13:17
Sample: 2010 2019
Periods included: 10
Cross-sections included: 20
Total panel (balanced) observations: 200
Swamy and Arora estimator of component variances

| Variable | Coefficient | Std. Error | t-Statistic | Prob. |
|---|---|---|---|---|
| C | 4104.845 | 1087.213 | 3.775565 | 0.0002 |
| YEBC | 0.363505 | 0.043412 | 8.373354 | 0.0000 |

Effects Specification

| | | S.D. | Rho |
|---|---|---|---|
| Cross-section random | | 4645.914 | 0.6531 |
| Idiosyncratic random | | 3386.093 | 0.3469 |

Weighted Statistics

| | | | |
|---|---|---|---|
| R-squared | 0.262259 | Mean dependent var | 1322.949 |
| Adjusted R-squared | 0.258533 | S.D. dependent var | 3924.703 |
| S.E. of regression | 3379.502 | Sum squared resid | 2.26E+09 |
| F-statistic | 70.38682 | Durbin-Watson stat | 0.336428 |
| Prob(F-statistic) | 0.000000 | | |

Unweighted Statistics

| | | | |
|---|---|---|---|
| R-squared | 0.327332 | Mean dependent var | 5890.529 |
| Sum squared resid | 6.19E+09 | Durbin-Watson stat | 0.122891 |

**Results of Hausman Test of Employment benefit cost and profit after tax**

Correlated Random Effects - Hausman Test
Equation: Untitled
Test cross-section random effects

| Test Summary | Chi-Sq. Statistic | Chi-Sq. d.f. | Prob. |
|---|---|---|---|
| Cross-section random | 0.229883 | 1 | 0.6316 |

Cross-section random effects test comparisons:

| Variable | Fixed | Random | Var(Diff.) | Prob. |
|---|---|---|---|---|
| YEBC | 0.356183 | 0.363505 | 0.000233 | 0.6316 |

Cross-section random effects test equation:
Dependent Variable: X1PAT
Method: Panel Least Squares
Date: 03/12/20   Time: 13:18
Sample: 2010 2019
Periods included: 10
Cross-sections included: 20
Total panel (balanced) observations: 200

| Variable | Coefficient | Std. Error | t-Statistic | Prob. |
|---|---|---|---|---|
| C | 4140.810 | 329.2926 | 12.57486 | 0.0000 |
| YEBC | 0.356183 | 0.046019 | 7.739888 | 0.0000 |

Effects Specification

Cross-section fixed (dummy variables)

| | | | |
|---|---|---|---|
| R-squared | 0.776998 | Mean dependent var | 5890.529 |
| Adjusted R-squared | 0.752081 | S.D. dependent var | 6800.554 |
| S.E. of regression | 3386.093 | Akaike info criterion | 19.19181 |
| Sum squared resid | 2.05E+09 | Schwarz criterion | 19.53813 |
| Log likelihood | -1898.181 | Hannan-Quinn criter. | 19.33196 |
| F-statistic | 31.18412 | Durbin-Watson stat | 0.370544 |
| Prob(F-statistic) | 0.000000 | | |

**Hypothesis for accepting the model:**

**Null hypothesis (H0)**- Random effects model is appropriate
**Alternative hypothesis (HA)**- Fixed effects model is appropriate

The probability value here is 0.6316 which is more than 5% thereby we accept null hypothesis and conclude that the random effect model is appropriate.

This result shows that random effect model is appropriate model to test the relationship between Employment benefit cost and profit after tax.

According to random effect model the probability value is 0.0002 which is less than 5%. So here by we reject H0 and accept H1, which means there is a strong relationship between Employment benefit cost and Profit after Tax.

# 6. CONCLUSION

The main aim of the study is to know the impact of Employment Benefit Cost on firm's profitability. Since the disclosures of Employment Benefit Cost are voluntary, there is a diversity of reporting practice. Large companies tend to report more Employment Benefit Cost in their annual reports than the medium-scale businesses; and the disclosure, tend to be more qualitative than quantitative despite the fact that there is a significant relationship between Employment Benefit Cost and Firm Profitability. The study shows the impact of Employment Benefit Cost on Firms profitability, there is a significant relationship between Employment Benefit Cost and Total assets, return on assets, Debt equity ratio, profit after tax and return on equity. The corporation organization should attach information about the value of Human Resource and the result of their performance during their accounting year in notes and schedule

# 7. LIMITATIONS OF THE STUDY

Future study can be done by taking more variables like market capitalization, return on capital employed, return on net worth etc. The current study is conducted on profitability of companies listed in National Stock exchange and further study can be conducted on companies listed in Bombay stock exchange, with reference to any specific sectors such as automobile sectors, pharmaceutical sectors etc.